\documentclass[graybox]{svmult}

\usepackage{mathptmx}       
\usepackage{helvet}         
\usepackage{courier}        
\usepackage{type1cm}        
\usepackage{makeidx}         
\usepackage{graphicx}        
\usepackage{multicol}        
\usepackage[bottom]{footmisc}

\makeindex             

\begin{document}

\title*{Alpha-cluster Condensations in Nuclei and Experimental Approaches for their Studies}
\author{Wolfram von Oertzen}
\institute{Wolfram von Oertzen \at Freie Universitaet, Fachbereich Physik and \\
Helmholtz-Zentrum Berlin, 
D14109 Berlin,  \email{oertzen@helmholtz-berlin.de}}
%
%
\maketitle

\abstract{The formation of alpha-clusters in nuclei close to the decay thresholds is discussed. 
These states can be considered to be boson-condensates, which are formed in a second order phase
 transition in a mixture of nucleons and $\alpha$-particles. The de Broglie wavelength of 
the $\alpha$-particles is larger than the
 nuclear diameter, therefore the coherent properties of the $\alpha$-particles give particular 
effects for the study of such states. The states are above the thresholds thus the enhanced 
emission of multiple-$\alpha$s into
the same direction is observed. The probability for the emission of multiple-$\alpha$s is not 
described by Hauser-Feshbach theory for compound nucleus decay.}

\section{Binding energy of alpha particles}
\label{sec:1}
The binding energies of nuclei in their ground states as a function of mass number 
show a peculiar systematic behavior, explained by the liquid drop model. 
Deviations from a smooth curve are due to shell effects, and are some times 
 discussed to be related to the formation of 
$\alpha$-clusters. The specific
properties of the nucleon-nucleon force, namely  the saturation which
 occurs if the spin and isospin
quantum numbers are both coupled to zero, produces  a very strong
binding of $\alpha$-particles~\cite{bm}. In addition, due to the 
internal structure of the $\alpha$'s an increased (30\%) central density
 is observed compared to the
the usual central density in nuclei. The $\alpha$-particle is therefore
a unique cluster subsystem in nuclei. 

This feature is well known from the early history of nuclear science,  and there has 
been small but steady activity in the field of clustering in nuclei in the last decades.
Recently more attention to clustering in nuclei has emerged due to the study 
of weakly bound nuclei at the drip lines. For these nuclei 
 clustering is very important even for the properties of ground states. These are 
 well reproduced in model independent approaches, like in the antisymmetrized 
Fermionic molecular dynamics (FMD) which uses  all degrees of freedom in the nuclear forces, 
in the approach by Feldmeier {\em et al.}~\cite{Feldmeier95,Neff05}. With 
a related approach, the 
antisymmetrized molecular dynamics (AMD)  with effective N-N forces, Horiuchi and
Kanada-En'yo {\em et al.}~\cite{horiuchi86,Hor97,Kan01} are able to reproduce
 the ground state properties and a large variety of excited  nuclear states,
 in particular those with molecular structure. 
In these calculations the density distributions of the nucleons are obtained. 
Quite spectacular are the results for 
loosely bound nuclear systems ~\cite{Hor97,Kan01}, where $\alpha$-clusters 
appear naturally as dominant substructures. 
This work has established that $\alpha$-clusters play a decisive role in the
description of light nuclei, in particular for the loosely bound neutron-rich
isotopes. For example the extra neutrons are found in covalent molecular orbitals around 
two $\alpha$-particles forming  bound molecular two-center systems for the
beryllium isotopes~\cite{voe06}.

Furthermore, the $\alpha$-particle is the most important ingredient in the concept 
of the Ikeda-diagram ~\cite{horiuchi68,Horiuchi70,Ikeda72}, where 
highly clustered states ({\em e.g.} linear chains) are predicted at
excitation energies around the energy thresholds for the decomposition
into specific  cluster channels.

\begin{figure*}[htbp]
  \begin{center}
    \includegraphics[width=0.85\textwidth]{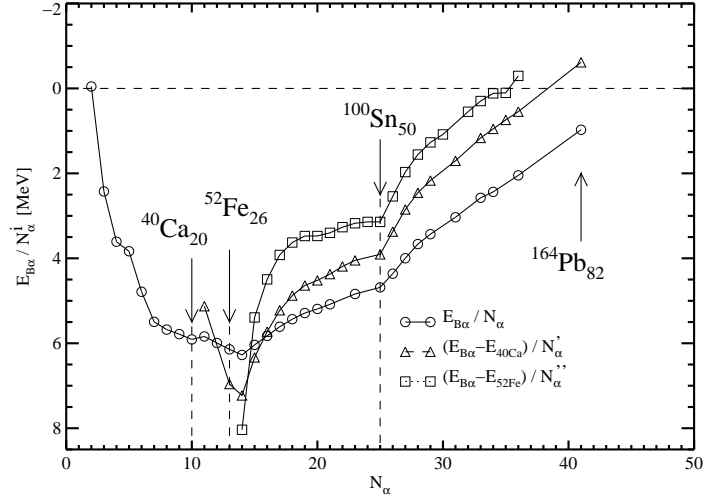}
    \vspace{-0.2cm}
    \caption{The experimental binding energies per $\alpha$-particle in N=Z nuclei,
             as function of the number of  $\alpha$-particles, $N_{\alpha}$.
             The lines are drawn to connect the points.
             The same quantities are shown under the assumption
             with two different heavy clusters as cores: $^{40}$Ca 
             and $^{52}$Fe, as indicated (adopted from~\cite{vOe06}) .}
    \label{fig:EB_alpha}
  \end{center}
\end{figure*}

In order to explore  the dynamics of $\alpha$-clustering in excited 
states of N=Z nuclei  the systematics of binding energies per
$\alpha$-particle in nuclei $E_{B{\alpha}}/N_{\alpha}$, has been
 considered~\cite{vOe06}. The experimental masses have been
taken from ref.~\cite{audi03}, for  $^{164}$Pb from a theoretical study of the mass 
A=164 region~\cite{isakov00}. 
With the total binding energy $E_B^{t}(N,Z)$, the binding energy of all $\alpha$-particles 
in  N=Z nuclei can be obtained, this will again 
follow the well known curve for the binding energy (per nucleon)
properly rescaled.
We are interested in the binding energy per
$\alpha$-particle, $E_{B\alpha}/N_{\alpha}$, determined from the experimental data,
 as  shown in Fig.~\ref{fig:EB_alpha}, there we show the quantity,
\begin{equation}
E_{B\alpha}/N_{\alpha} = 
[E_B^{t}(N,Z) - N_{\alpha}E^{\alpha}_B]/N_{\alpha}.
\end{equation}
Here we have the typical maximum values around Fe-nuclei. The 
energy for the threshold states can be read from this figure.
Searching for the alpha-condensed states we can also include states 
with a strongly bound core, e.g. a core of  $^{16}$O, a core of $^{40}$Ca  
or $^{52}$Fe, with the appropriate  number  $N^{'}_{\alpha}$ of free $\alpha$-particles outside. 
The binding energy with a core shown in  Fig.~\ref{fig:EB_alpha} being:
\begin{eqnarray}
\lefteqn{E_{B{\alpha}}^{^{40}Ca}(N^{'}_{\alpha})/(N_{\alpha}) = }\\
& \left[ E^{t}_B(N,Z)- E_B^{^{40}Ca}(N,Z) - (N_{\alpha} - 10) E^{\alpha}_B
\right]/(N_{\alpha}-10). \nonumber
\end{eqnarray}
\noindent
From this figure we can deduce that  the excitation energy, where the value 
of $E_{B\alpha}/N_{\alpha}$ reaches zero, 
the threshold for complete decay, becomes lower with the inclusion of a core.
 For heavier nuclei around  $^{100}$Sn and masses 
approaching  $^{164}$Pb the 
nuclei become unstable relative to single (or multiple) proton or  $\alpha$-particle emission.
The excitation energies, where the binding energy value for  $\alpha$'s approaches zero
- these are the values where  $\alpha$-particle condensates can form - 
will be discussed below, sect.~\ref{sec:cond}. Although these $\alpha$-condensed
 states are at rather 
 high excitation energies in the continuum of nucleonic states, 
they may have collective properties, which can give them a smaller observable width. 
We expect the decay into many  $\alpha$-particles, a decay not 
described by the Hauser-Fehsbach formalism for statistical 
compound nucleus decay (see below).

\section{The formation of alpha condensates }
\label{sec:cond}
The $\alpha$-particle condensates formed at the thresholds will be unbound states, 
their decay properties will be one of the
most important points in the discussion of these boson states. In lighter nuclei, 
in particular for the second
 $0_2^+$ state  in $^{12}$C, the {\em Hoyle state}~\cite{hoyle} ,
 which can be considered as the first boson condensate, 
a gamma-decay is possible, here with  
 the sequence  $0_2^+$ $ {\rightarrow}$  $2_1^+$ $ {\rightarrow}$ $0_1^+$, a process most important for the
 formation of $^{12}$C in stars. The recent proposal of an  $\alpha$-particle 
condensate wave function (THSR), by  Tohsaki, Horiuchi, Schuck and R\"opke 
 ~\cite{tohsaki01}, describes the properties of the Hoyle state very well, whereas
even the largest shell model calculations fail completely to
 reproduce this state~\cite{schuck04a,schuck04b}, see also refs.~\cite{Schuck07}
for the most recent discussion.

 In  medium size nuclei (Z$<$20) the $\alpha$-condensates,  calculated using the known 
alpha-alpha-potential  with a
 self-consistent approach (based on the Gross-Pitaevskii equation), will have  Coulomb
barriers~\cite{yamada04} for the decay in to multiple  $\alpha$'s.
 With these barriers the states will have  sufficiently small width
 for potential studies by inelastic scattering.
 However, the heaviest nucleus for which this
barrier can create a quasi-bound state~\cite{yamada04}, is estimated to be around 
 $^{40}$Ca. In heavier nuclei 
these states will be embedded
 high in the continuum of the fermionic states,
their decay is expected to be non statistical,  the most characteristic  property to study.

\subsection{Second order Phase transition}
\label{subsec:2}
The ground states of nuclei are well described by the shell model
 with a self-consistent
potential of all nucleons. If a cluster model with  $\alpha$-clusters 
is used,  their strong spatial overlap,
  the anti-symmetrization of all nucleons destroys  
their original properties, a fact widely discussed in the literature 
(see refs.~\cite{funaki09} for the most recent discussion). 
 The intrinsic structure of 
$\alpha$-clusters in these cases are very different from that of 
free $\alpha$-particles, still a large variety of molecular resonances 
connected to clusters 
are observed in  N=Z nuclei~\cite{freer07}.               
 
We want to discuss the formation of an  $\alpha$-particle gas,
 where the average distance between $\alpha$-particles is much larger 
with rather small spatial overlap. The 
 corresponding nucleon density will be well below normal nuclear densities.
In fact, in the theoretical investigation of Bose-Einstein condensates in nuclei
 Tohsaki, Schuck {\em et al.} ~\cite{tohsaki01,schuck04a,schuck04b,yamada04}
 find, that at the thresholds for multi $\alpha$-particle decays,
the states with $\alpha$-clusters have a much larger radial extension than
the ground states (larger  $\alpha$-$\alpha$ distances).
From the view point of the nucleonic fermion gas the  appearance of such states will 
depend on the temperature 
({\em i.e.} excitation energy, $E_{x}^{\ast}$) of the nucleus.  The
concept of a second order phase transition as in a chemical reaction  with two components 
 can be used~\cite{vOe06},  a concept well 
established in thermodynamics of composite systems in 
statistical physics ~\cite{amit}.
The basic equation is the ``reaction'' of four ``free'' nucleons 
(two protons and two neutrons  coupled to total values of spin 
and isospin of zero)  forming $\alpha$-clusters:\\
$(N_1 + N_2 + N_3 + N_4)\ {\longleftrightarrow}\ {\alpha}$-particle
+ 28.3 MeV.\\
\noindent
The free nucleons, $N_i$, should have a definite volume and pressure, 
in order to define thermodynamic quantities and where the density allows
the occurrence of the mentioned reaction. We can assume that the particles interact in a 
well defined volume created by 
 a self consistent mean field for the nucleons (the Hartree-Fock approach) and
 for the $\alpha$-clusters with the Gross-Pitaevskii approach for bosons. This latter 
 has been used in  the work of Yamada and Schuck~\cite{yamada04}. 

\begin{figure}[htbp]
  \begin{center}
    \vspace {-1.0 mm}
    \includegraphics[width=0.69\textwidth]{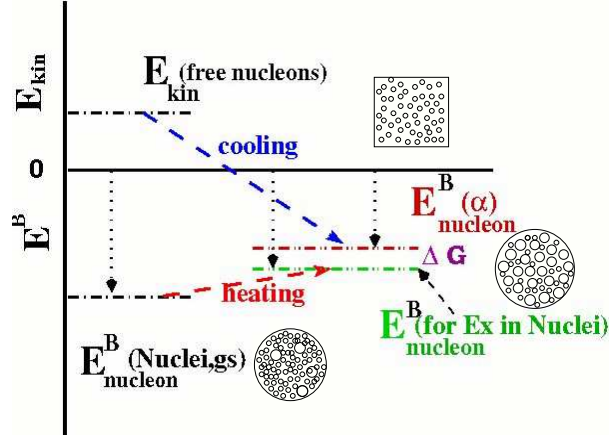}
        \vspace {1.0 mm}
    \caption{Schematic illustration of the relative values of the energies of
             free nucleons and alternatively their binding energies  
             in nuclei (8.2 MeV), the latter are generally larger than 
             in $\alpha$-clusters (7.07 MeV). The difference $\Delta$G 
             between these two 
             binding energies decreases 
             with increasing excitation energy (``heating'')  in nuclei. At a  critical value
             the binding energies become equal, ${\Delta}G=0$,
             a collective state of bosons (potentially mixed with fermions), 
             the condensed $\alpha$-particle gas
             can be formed.
             In the AMD this state is approached by ``cooling'' from a fermion gas.}
    \label{fig:EbindEx}
  \end{center}
\end{figure}

 In models like the AMD ~\cite{Kan01} a certain number of 
nucleons are confined in a volume with a positive kinetic energy, as 
suggested in Fig.~\ref{fig:EbindEx}. In this model a cooling method is applied
to find the states of the lowest energy and higher density.  The energy of the
nucleons inside the nucleus is defined by their volume, and their
Fermi-energy can be deduced 
from the nuclear radius, as described in text books~\cite{FH}. In the AMD approach 
with the cooling process 
a certain $\alpha$-cluster phase is observed, before the formation of the  higher density states 
with bound fermions, and finally the ground states are reproduced. 
At the end the formation of normal nuclei 
with a binding energy per nucleon of 8.2 MeV or more is observed, a value which is 
higher than in the  $\alpha$-cluster (7.073 MeV). These two values define the difference
in the chemical potentials in the two phases  (Fig.~\ref{fig:EbindEx}).
 For less bound nuclei (binding energy per nucleon around 7.073 MeV), the $\alpha$-clusters
are obtained in a ``natural'' way.
Starting from the ground states of normal nuclei the nucleons will  form  an $\alpha$-cluster 
phase, with increasing temperature of the nucleus, {\em e.g.} with 
increasing excitation  energy, see  Fig.~\ref{fig:EbindEx}, (``heating''). This excitation
 energy becomes rather
low in neutron-rich light exotic nuclei, where clustering may appear already
 in the ground states as the
dominant structure~\cite{Kan01}, this may also happen potentially
 in very heavy N=Z nuclei.

For the nucleons confined in the nuclear volume we apply the concepts
of statistical physics for the reaction $4N$ ${\longleftrightarrow}$ $\alpha$-particle.
The rate of the reaction is governed by the free energy, $G$,
and the difference in the  chemical potentials, $\mu_{\alpha}$ and $\mu_{n}$.
The chemical potentials are defined as 
$\mu_{i}$ = ${\delta}G/{\delta}N_i$, $i=n,{\alpha}$.
The thermodynamic free energy depends on the number of nucleons,
$N_{n}$ and on $N_{\alpha}$, with $G = G(N_{n},N_{\alpha}$).
The change of the free energy becomes
\begin{equation}
\Delta G = \Delta N_{\alpha}\mu_{\alpha} + 4 \Delta N_{n}\mu_{n}.
\end{equation}
For the phase transition a minimum value 
of the free energy is needed, this gives the condition ${\Delta}G=0$, this
feature will be observed at a critical excitation energy $E^{crit}_{x}$.
In the nuclear medium $\Delta G$ is the difference between the binding energy
of the four  nucleons in the free  $\alpha$-particle to that in the nuclear medium,
as illustrated in Fig.~\ref{fig:EbindEx}.

The kinetic energy of the nucleons determines the temperature, $T$.
However, we will use the temperature of the nucleus, $t$, related to its
excitation energy. In the normal case of a mixed system of the two species,
the relative abundance of $N_{\alpha}$ to $N_{n}$
is a function of the temperature (in our case excitation energy) 
and is obtained through the expression 
\begin{equation}
\frac{N_{\alpha}}{(N_n)^4} = K(t) = \exp(-\frac{\Delta G(t)}{RT(t)})
\end{equation}
\noindent
The  value of $K$ is to be determined by experimental observation 
(the usual coefficient $R$  appears as in statistical physics).
For the case of \emph{negative} ${\Delta}G(t)$, a decrease of the 
free energy (corresponding to a large value of the ratio $K$)
gives a higher density of the $\alpha$-particles as reaction products.
A \emph{positive} value of  ${\Delta}G$ corresponds to an energetic disadvantage
for the reaction creating  $\alpha$-particles, resulting in a smaller number $N_{\alpha}$
 as  reaction products.
In the case of nuclei, the nucleons are embedded in the nuclear medium and are
confined in the nuclear potential created by the mean field of all nucleons.
The binding energy per nucleon in nuclei is around 8 MeV 
(dependent on the size of the nucleus and its excitation energy).
The nucleons  have a larger binding energy in the nuclear medium (in the ground states
of stable nuclei) compared to the value in the $\alpha$-clusters.
The relative positions of the relevant energies are illustrated in
Fig.~\ref{fig:EbindEx}, from ref.~\cite{voe06}.
 Actually, because the chemical potential
of the nucleons will depend on the excitation energy in the nucleus
(or on its temperature), we put this dependence in the expression for 
${\Delta G(t)}$. 

\begin{table}[htbp]
\addtolength{\tabcolsep}{-0.3mm}
\caption{Alpha-particle binding and critical excitation energies for the 
         condensation condition in nuclei with N=Z;
         $N_{\alpha}$ - number of $\alpha$-particles,
         $E^t_B/{N_n}$ - binding energy per nucleon, 
         $E_{B\alpha}/N_{\alpha}$ - binding energy per $\alpha$-particle,
         $E^{crit}_{x}$ - condensation energy. The last column shows 
         the values for the case of a $^{40}$Ca-cluster core.
         All energies in MeV.}
\begin{center}
\begin{tabular}{|l|l|l|l|l|l|r|}
\hline
Nuclide & $N_{\alpha}$ & $E^{t}_{B}$ & $E^{t}_{B}/{N_n}$ & $E_{B\alpha}/N_{\alpha}$
& $E^{crit}_{x}$ &  $E^{crit}_{x}$ \\
\hline
$^{4}$He    &  1    & 28.3   & 7.073    & ---      & - & ($^{40}$Ca)\\
\hline
$^{12}$C    &  3    &  92.16 & 7.680    & 2.425  &   7.27   & ---   \\
$^{16}$O    &  4    & 127.6  & 7.976    & 3.609  &  14.44   & ---   \\
$^{20}$Ne   &  5    & 160.7  & 8.032    & 3.83   &  19.17   & ---   \\
$^{24}$Mg   &  6    & 197.2  & 8.260    & 4.787  &  28.72   & ---   \\
$^{28}$Si   &  7    & 236.5  & 8.447    & 5.495  &  38.47   & ---   \\
$^{32}$S    &  8    & 271.8  & 8.493    & 5.677  &  45.41   & ---   \\
$^{36}$Ar   &  9    & 306.7  & 8.519    & 5.78   &  52.02   & ---   \\
$^{40}$Ca   & 10    & 342.0  & 8.551    & 5.910  &  59.10   & ---   \\
$^{52}$Fe   & 13    & 447.7  & 8.609    & 6.143  &  79.86   & ---   \\
$^{56}$Ni   & 14    & 483.9  & 8.642    & 6.275  &  87.85   & ---   \\
$^{72}$Kr   & 18    & 607.1  & 8.432    & 5.433  &  97.8    & 87.78 \\
$^{80}$Zr   & 20    & 669.8  & 8.371    & 5.192  & 103.8    & 90.38 \\
$^{100}$Sn  & 25    & 824.5  & 8.244    & 4.684  & 117.1    & 97.65 \\
$^{112}$Ba  & 28    & 894.8  & 7.99     & 3.665  & 102.6    & 68.79 \\
$^{144}$Hf  & 36    & 1090.9 & 7.577    & 2.074  &  74.6    & 19.68 \\
$^{164}$Pb  & 41    & 1200.1 & 7.317    & 0.973  &  39.9    &$-$25.21\\
\hline
\end{tabular}
\end{center}
\label{tab:ecrit}
\end{table}
Alpha-cluster formation is expected if $4E^t_B/{N_n}$ is less than or equal
to the total binding energy of four nucleons in the $\alpha$-cluster.
As the binding energy per nucleon becomes equal or smaller than in
the $\alpha$-particle, a new phase will be formed, a strongly
interacting Bose gas. For binding energies of the 
nucleons  close  to (or larger) that
in the $\alpha$-particle it becomes possible to form 
a mixed phase of $\alpha$-cluster states (liquid) and of nucleons.
The binding energy of nucleons in the ground states 
of nuclei (see Fig.~\ref{fig:EbindEx}) 
$E^t_B/{N_n}= E^B_{nucleon}$, 
is usually larger than in the $\alpha$-particle.
The condition for the excitation (condensation) energies is $E_{x}^{cond}{\ge}E^{crit}_{x}$. 
The values for different nuclei relevant to this concept
are given in Tab.~\ref{tab:ecrit}.

We sumarize that the $\alpha$-condensation condition is given by \\
$E^t_{B}/{N_n}(E^{crit}_{x}){\ge}7.07$ MeV, \\
The value of the critical excitation energy/(per nucleon)
 in a nucleus, $E^{crit}_{x}$, should be equal or larger  than 7.07 MeV,
 which is the binding energy
of nucleons in the $\alpha$-particle. This statement is the same 
as the condition $\Delta G(t)=0$. Alternatively, the phase transition will be achieved 
at excitation energies of the nucleus,
$E_{x}^{\ast}$, corresponding to the thresholds where  all clusters become
unbound, the condition being that  $E_{B\alpha}(N,Z)$ = 0. This is the 
original concept of the Ikeda diagram.
The  Ikeda diagram~\cite{horiuchi68} gives a phenomenological
condition for the appearance of clustered states (with the inclusion of
other clusters like  $^{12}$C, $^{16}$O, {\em etc.}) in nuclei.
 We can state that the Ikeda diagram  with $\alpha$-particles can be  
deduced from  thermodynamic considerations.
The level density for the fermionic phase space grows very fast
 with excitation energies, whereas those for the bosons will grow much slower.
\begin{figure}[htbp]
  \begin{center}
    \vspace {2.0 mm}
    \includegraphics[width=0.98\textwidth]{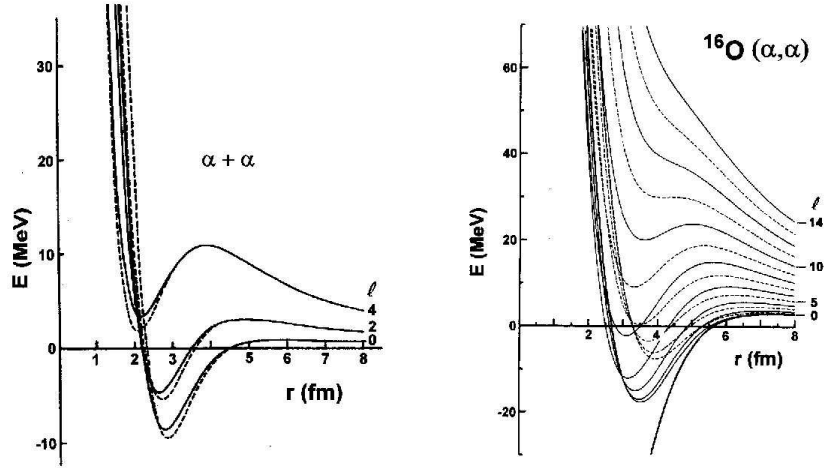}
        \vspace {2.0 mm}
    \caption{The equivalent local potentials describing the  $\alpha$-$\alpha$ and
             the $\alpha$~-~$^{16}$O interaction.
             The resonant energies in $^8$Be and the phase shifts are reproduced. 
             With these interactions  the $\alpha$-particle gas can be calculated.
             For the  $\alpha$~-~$^{16}$O potential a similar van der Waals type of interaction
             is  obtained, a potential suited for an  $\alpha$-particle gas with 
              additional binding by a core (adopted from ~\cite{voe06}).}
    \label{fig:a-a-pot}
  \end{center}
\end{figure}

Most important for the properties of the  $\alpha$-particle gas 
is, that they do not represent the ``ideal'' gas, they interact via an interaction
which has similarities with a van der Waals interaction, with a strongly repulsive core due to 
the Pauli principle, see Fig.\ref{fig:a-a-pot}.
Two $\alpha$-particles 
form as the lowest state, the ground state of $^8$Be, a resonance
at $E^{\ast}_{x}=92$ keV. We can calculate the de Broglie wave length, 
$\lambda =  {h}/{\sqrt{(2 \mu E^{\ast}_{x})}}$  for this case
and have $\lambda$ = 67 fm (relative motion between the two  $\alpha$-particles).
 If for higher excitation we incorporate
the 2$^+$ at 3.04 MeV the value of $\lambda$ is still 12.4 fm.
Similarly three $\alpha$-particles can form the Hoyle-state just 
above the three $\alpha$-particle threshold in 
$^{12}$C, the 0$^+$ at 7.654 MeV (288 keV above the threshold of 7.346  MeV). 
With these values for three $\alpha$-particles
we again get a similarly large de Broglie wave length of relative motion.
 Also the third 0$^+_3$
at 10.3 MeV excitation energy 
can participate in the formation of a multi-$\alpha$-particle correlation.

 Overall we have values for $\lambda$ in the condensed state larger (by factors 2-5)
 then the radial extension of the nucleus. The  multi-$\alpha$-particle states will
contain the  $\alpha$-particles mainly in their resonant states  in 
$^8$Be. The condensed states at the binding energy threshold 
consisting of $\alpha$-particles will form coherent super-fluid states. 
The resonant states in $^8$Be and $^{12}$C act in a similar way as the 
residual interaction in the formation of the superfluid neutron pairing 
states, see ref.~\cite{bm}, volume II. 
The calculations of  THSR  based on a local $\alpha$-$\alpha$ 
potential reproduce the states of  $^8$Be, and the threshold states in  other light
 nuclei.
Inspecting the local potentials in Fig.~\ref{fig:a-a-pot} for the system of 
$^{16}$O+$\alpha$-particle,
we conclude that alpha-condensates with a $^{16}$O-core can be formed, where this potential
will create a common binding potential, for a larger number of alpha's 
(e.g. $^{40}$Ca = $^{16}$O+6$\alpha$).
\begin{figure}[htbp]
  \begin{center}
  \vspace {-0.0 mm}
    \includegraphics[width=0.78\textwidth]{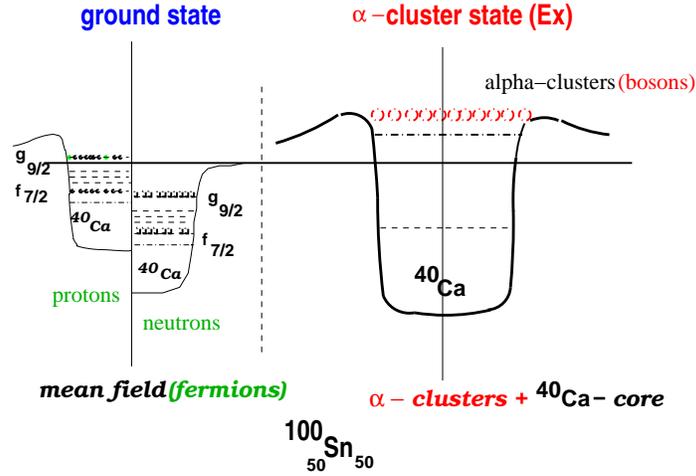}
    \vspace{0.1cm}
    \caption{Schematic illustration of the two models 
             for states in $^{100}$Sn.
             States of low excitation energy are formed by the mean
             field of nucleons, in this case the potentials for neutrons and protons 
             are rather different due to the Coulomb interaction. Thus, the 
             formation of $\alpha$-particle structures is 
             strongly suppressed. At the critical excitation energy of 97 MeV
             (for $^{100}$Sn, see Fig.~\ref{fig:EB_alpha}),
              a collective state of bosons with $\alpha$-particles occupying 
             the same orbit (relative S-states) outside a $^{40}$Ca-core, 
             will be energetically favored.}
    \label{fig:100Sn}
  \end{center}
\end{figure}

In Fig.~\ref{fig:100Sn} we illustrate the possible situation for an 
 $\alpha$-condensate in  $^{100}$Sn,
with a core of  $^{40}$Ca and 15 $\alpha$'s. These configurations 
 can be formed in a reaction with a $^{72}$Kr beam and a $^{28}$Si target.
At excitation energies of 97 MeV or more (excitation energies discussed earlier)
many compound nuclear (CN) states will exist, consisting of different
configurations of the $\alpha$-particle gas plus a core. Here again the 
threshold rules apply with respect to excitation energies. 
We may expect many overlapping states 
(with a large decay width), which will interact 
coherently (see ref.\cite{Tza06a}), because 
the same compound states of the $\alpha$-particle phase can be formed  
with a different number of 
 $\alpha$-particles. These
will interact through the 0$^{+}$ and 2$^{+}$ resonances of $^{8}$Be
and $^{12}$C$^*$, depending on 
the excitation energy of the state. The decay of such a state (in the figure  
 there is no barrier for the  $\alpha$-particles, in difference to the figure
 shown in ref.~\cite{voe06}) can occur sequentially with different energies in each step, 
as in CN-decay. However, 
the most interesting case would be the simultaneous decay, with many  $\alpha$-particles 
with almost equal kinetic energies, a process, which can also
 be considered as Coulomb explosion~\cite{jortner},
see sect.\ref{exper:3}.

\section{Experimental observables}
For the observation of states in nuclei, which have spin(parity) = 0(+) and
 the properties of 
  $\alpha$-particle condensates, there are several characteristic
experimental features which we can propose for future studies.\\
1) The study of the radial extension, e.g. observed in inelastic
 $\alpha$-scattering and in
the form factors from electron scattering experiments.\\
2) Coherent emission of $\alpha$-particles from compound nuclei
in coincidence with large $\gamma$-detection arrays.\\
3) Fragmentation into multiple $\alpha$-particle channels at GeV/nucleon energies.\\
4) $\alpha$-$\alpha$-correlations, for CN decay similar to 2.\\

A further approach which should be mentioned here is the detection
 of multi-$\alpha$-clusters in 
a ternary cluster decay as described in refs.~\cite{Zhereb07,VOe08}. In these cases
the coplanar detection of two heavier fragments as in a binary decay,
 shows missing mass and charge of multi-$\alpha$-clusters. 
The study of these fission processes  indicates that
the missing $\alpha$-clusters are emitted from the neck 
with very small intrinsic excitation and small angular momentum.
 These  multi-$\alpha$-clusters,
will  be emitted towards very small angles, where they should be detected with
charged particle counter-telescopes.

\subsection{Inelastic scattering, radial extensions,
 form factors}
\label{exper:1}
The threshold states in nuclei with condensed $\alpha$-particles have
 spin/parity $J^{\pi} = 0^+$, these 
must be populated by  monopole excitations (a collective radial density mode).
 In fact the most
 important predicted  properties of 
these states, are the larger 
radial extensions. These can be manifested in inelastic
 electron and hadron scattering.
 The inelastic electron scattering on  a $^{12}$C-target has been studied
repeatedly. The form factor for the transition to the excited state
 at 7.65 MeV, the $\emph {Hoyle State}$, with 
$J^{\pi} = 0^+$  is thus well known see refs.~\cite{Chern07,Khoa08}
 and earlier references therein.
\begin{figure}[htbp]
  \begin{center}
    \includegraphics[width=0.80\textwidth]{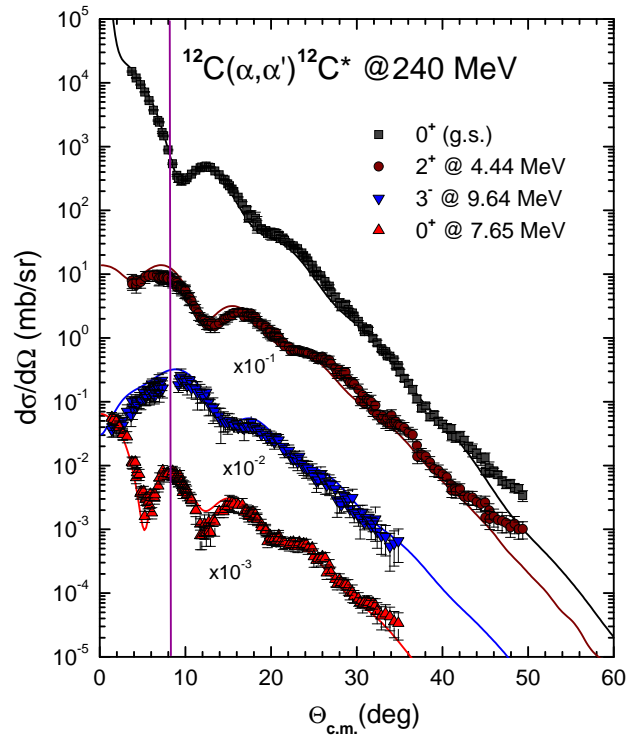}
        \vspace {-4.0 mm}
    \caption{The elastic and inelastic scattering of  $\alpha$ + $^{12}$C at 240 MeV with 
             the result of the analysis using the double folding 
             model of Khoa et al.~\cite{Khoa08}.
             The vertical line illustrates the shift towards smaller angles for the transition
             to the  
               7.65 MeV, the ($\emph {Hoyle state}$), due to it's larger radial extension.}            
    \label{fig:a-12C}
  \end{center}
\end{figure}

 Similarly there are extensive studies of inelastic hadron  
scattering on  $^{12}$C using a large variety of projectiles. We concentrate here on
 the elastic and inelastic $\alpha$-scattering 
at energies between 104 MeV and 240 MeV, which has recently been analyzed with microscopic 
transition densities and the double folding
approach for the scattering potential~\cite{Khoa07} and with a diffraction
 model~\cite{Danilov09}. The angular distributions exhibit 
at smaller angles strong diffraction patterns, and partially also a refractive maximum 
at larger angles. One feature, known from the early
 history of nuclear physics, is the Blair phase-rule established in  $\alpha$-particle 
scattering~\cite{Khoa08}. If states are populated in inelastic scattering and 
sufficiently high energy, diffractive
 patterns (strong maxima and minima) 
are observed at forward angles, the structures of the elastic scattering (e.g. the minimum) 
and the  e.g. the maximum for (Fig.~\ref{fig:a-12C}) in  inelastic scattering  ($L=0$) are exactly 
 {\em out of phase if no parity change has occurred}. At higher 
energies the effects of Q-values, their influence on  the position of the 
maxima and minima is small. The position of the diffractive minima depend 
on the radial extension (e.g. of the excited states).

The result at an incident energy
 of 240 MeV  is shown in Fig.~\ref{fig:a-12C}, the angular distributions
 show pronounced diffraction structures. Indeed the diffractive pattern for the 
inelastic excitation to the 
 2$^{+}$ state at 4.43 MeV is clearly  out of phase with that for the ground state.
For the 0$^{+}$ state at 7.65 MeV the diffractive pattern is more
 pronounced and is 
  shifted by approximately 2 degrees to forward angles relative to the elastic
scattering, indicating a larger radius. 
The calculations, which were performed with
 the double folding model for the elastic scattering potential as well as for the
 transition densities~\cite{Khoa08},
are  also shown in Fig.~\ref{fig:a-12C}. With this approach and a proper
 choice of the imaginary potential for the 0$^{+}_2$ state 
the absolute cross sections are reproduced with a correct value
for the $E0$-transition strength.
 The 
 other inelastic transitions have been calculated, and are perfectly reproduced
due to the choice of the transition densities obtained in the folding model. 
The analysis  with a diffraction
 model~\cite{Danilov09} of such data gives the  systematics of the diffraction radius over
 a large energy range
and indicates a 10$\%$ larger radius for the 0$^{+}_2$ state compared to
 the ground state.
 Similar results will be expected
for the 0$^{+}_6$ at 15.1 MeV in  $^{16}$O, which is just above the
 4$\alpha$-threshold (14.4 MeV),
 and has been searched for recently~\cite{Wakasa07}.

\subsection{Compound nucleus decay, correlated emission 
of alpha's}
\label{exper:2}
In the formation of N=Z compound nuclei up to mass A = 60-80, the heaviest combination of 
stable targets and projectiles is $^{40}$Ca + $^{40}$Ca  giving 
$^{80}$Zr compound states with appropriate excitation energy (see 
Tab.~\ref{tab:ecrit}) and 
 coherent  $\alpha$-particle states can be formed.
For heavy N$>$Z compound nuclei with a small neutron excess, however, 
the features  discussed below may also apply.
 For even heavier systems with N=Z, 
we will have to resort to beams of unstable nuclei, like e.g.
a  $^{72}$Kr beam, which has a good chance of
being produced in the future with usable intensities.
The compound nucleus with a $^{40}$Ca-target will be $^{112}$Ba
($Q=-52.54$ MeV).
Because of the fact, that the heavier compound nuclei are very far off-stability, the 
reaction $Q$-value becomes very negative. With an incident energy close to the 
Coulomb barrier, the final excitation energy (Ex) can be well controlled and moderate 
values of Ex can be reached (see Tab.~\ref{tab:ecrit}).
These compound nuclei will have also favorable 
 $Q$-values for the emission of 
several $\alpha$-particles. Actually a new collective decay mode,
 where all alpha-particles share
 the same kinetic energy, as in Coulomb explosion, can be predicted. However,
 heavier compound nuclei will 
 be unstable to charged-particle emission (protons and  $\alpha$'s) 
already in their ground states.

 Further we may consider an excess of two or more neutrons (with an isotope with a more
intense beam), this would most likely not destroy the special states discussed here.
The excess neutrons will be placed in quantum orbits around the emitted  clusters,
for example as in the $^{9-10}$Be isotopes forming bound or metastable molecular states 
and configurations with low nucleon density~\cite{voe06}.

We are interested in the
 multiple $\alpha$-particle emission. Due to the coherent 
properties of the threshold states consisting of $\alpha$-particles
interacting coherently with a large de Broglie wave length,
the decay of the CN will not follow the Hauser-Feshbach assumption 
of the statistical model: that all decay steps are statistically independent.
If we consider a sequential process, 
after emission of the first $\alpha$-particle, the residual nucleus contains
the phase of the first emission process; the subsequent decays will follow 
with very short time delays related to nuclear reaction times 
(or their inverse, decays), favoring the formation of resonances like  
$^{8}$Be$(0^{+},2^+)$ and 
the $^{12}$C$^\star(0^{+}_2,2^{+}_2)$ states.

 Another view for the $\alpha$-gas in nuclei is the concept of 
 a collective super-fluid state
with a broken symmetry, the $\alpha$-particle number, a concept much used
for neutron pairing in superfluid states in nuclei~\cite{bm,broglia}. 
For the two-neutron pairing states in heavy nuclei, the 
transfer of neutron pairs between 
superfluid nuclei~\cite{broglia,voevitt01} is strongly enhanced. The 
analogy to the enhancement of the transfer of correlated neutron-pairs, 
is the multiple emission of $\alpha$-particles 
as a collective transition (changing the particle-number as a 
collective variable) from compound nuclei with 
superfluid properties (with  $\alpha$-condensates), i.e.
 between nuclei with different numbers of
$\alpha$-particles. This feature has been 
 discussed for $\alpha$-particle 
transfer between very heavy nuclei in the valley of stability in
ref. ~\cite{voe76}. Thus  
the observation of enhanced  multiple emissions of $\alpha$-particles 
from the compound state 
can be proposed as the signature for the observation 
of the collective Bose-gas. These multiple emission should be 
strongly enhanced relative 
to the statistical model prediction. For the latter case the emission 
of several $\alpha$-particles would be observed into different angles 
\cite{Tza05}.
\begin{figure}[htbp]
  \begin{center}
      \includegraphics[width=0.75\textwidth]{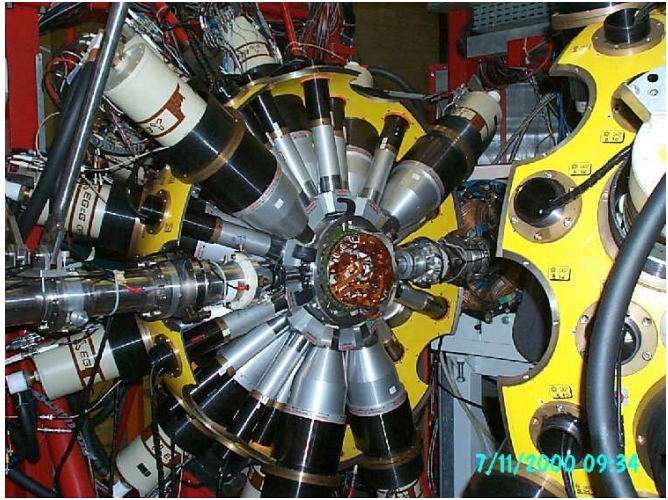}
     \caption {Picture of the $\gamma$-detector ball GASP, opened to give view on 
               the detector ball ISIS with 42 
               $\Delta$E-E telescopes placed inside. With these
               the emission of three 
             $\alpha$'s at different angles in different detectors 
            (upper panel in Fig.\ref{fig:gamc}),  
            or the pile-up events for the  3$\alpha$'s  from the decay of $^{12}$C$^*_{0^+}$,
               see  Fig.\ref{fig:dE_E_2},  are registered (courtesy of A. de Angelis, LNL).}
    \label{fig:gaspisis}
  \end{center}
\end{figure}

The coherent emission should occur into the same (identical) angle. This 
will lead to the situation  that the 
observation of unbound resonances becomes possible, such as  $^{8}$Be$(0^{+},2^+)$ 
and the excited states of   $^{12}$C, the
$^{12}$C$^\star(0^{+}_2 ,0^+_3)$-clusters. This  feature in fact
has been observed in the recent data 
~\cite{Torilov,Tza05,Tza06a,thummerer00,thummerer01,vOe00} discussed 
in the next section.

\subsection{Compound states with multi-$\alpha$ decays}
\label{exper:3}
We are interested in the
coherent multiple $\alpha$-particle emission from excited compound nuclei (CN).
 Due to the coherent 
properties of the threshold states consisting of $\alpha$-particles
interacting with a large de-Broglie wave length~\cite{voe06},
the decay of the CN will not follow the Hauser-Feshbach assumption 
of the statistical model: a sequential decay and 
that all decay steps are statistically independent.
\begin{figure*}[htbp]
 \begin{center} 
   \vspace {-2.0 mm}
   \includegraphics[width=0.75\textwidth]{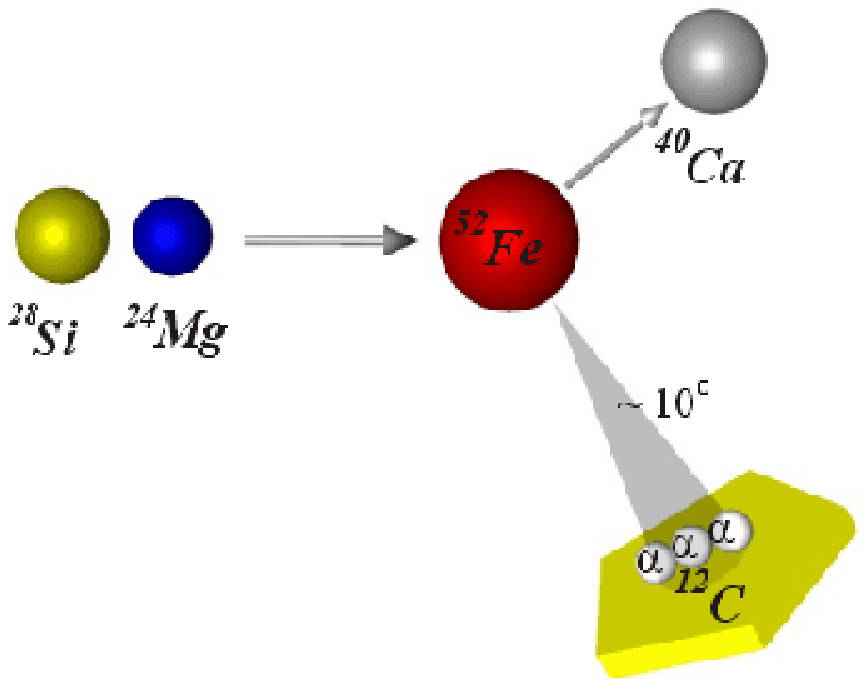}
      \vspace {2.0 mm}
    \includegraphics[width=0.60\textwidth]{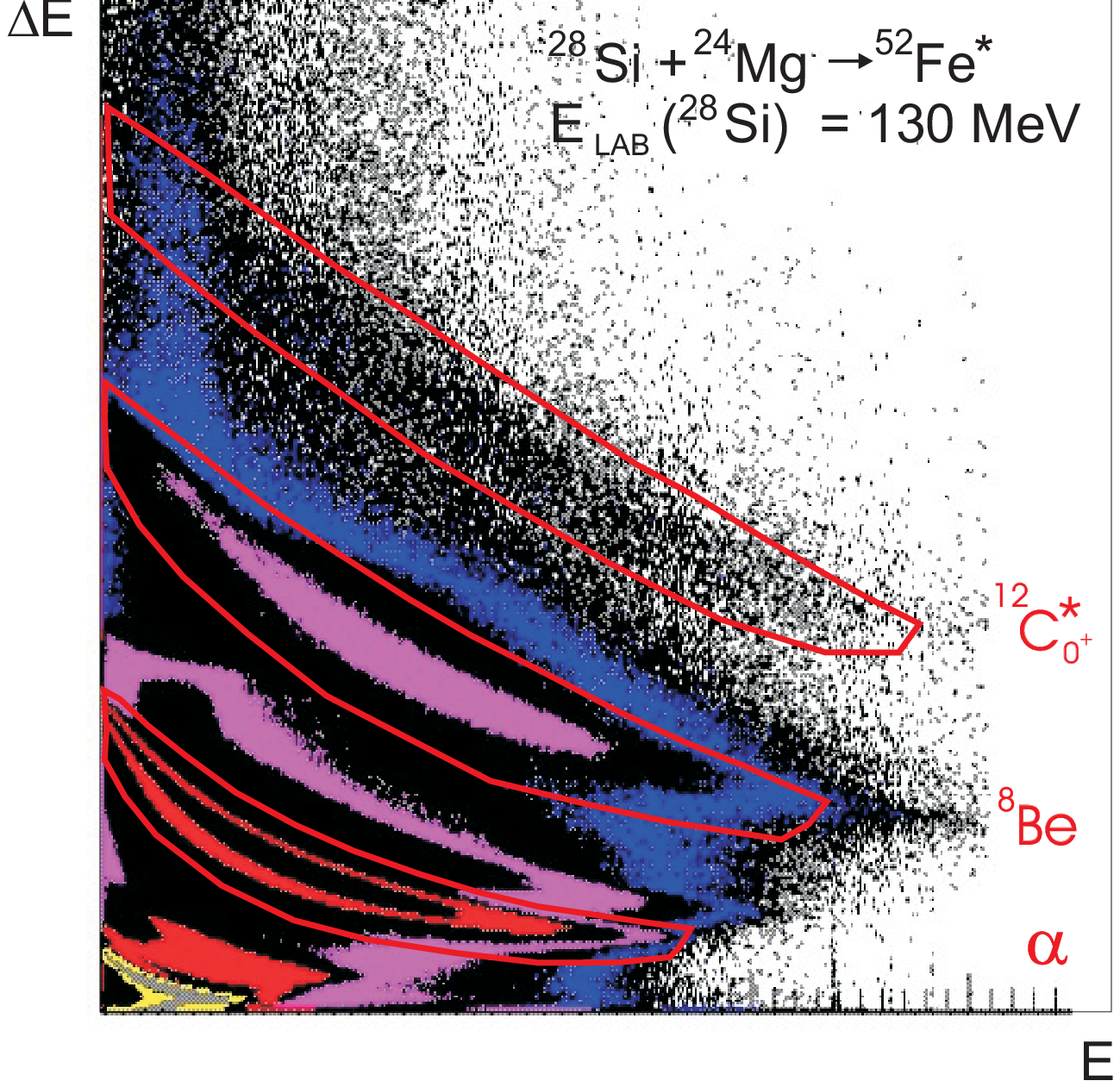}
     \caption {Top: This  part shows the kinematical situation for the triple
            pile-up of the signals for 3 $\alpha$'s in one detector and the
            emission cone for  $\alpha$'s from the decay of $^{12}$C$^\star(0^{+}_2)$. 
              Bottom: Plot of $\Delta$E-E-signals as observed with the 
            ISIS charged particle detector system. The events with the emission of
            single $\alpha$'s, of $^{8}$Be and with three  $\alpha$'s from
            the  state  $^{12}$C$^\star(0^{+}_2)$
            are indicated. 
            The reaction is  $^{28}\textrm{Si} + ^{24}\textrm{Mg} 
            \rightarrow  ^{52}\textrm{Fe}
            \rightarrow  ^{40}\textrm{Ca}+ X$ at E$_{lab}$=130 MeV (courtesy of Tz. Kokalova).}
    \label{fig:dE_E_2}
  \end{center}
\end{figure*}

After emission of the first $\alpha$-particle, the  residual $\alpha$-particles in the  
 nucleus contain
the phase of the first emission process; the subsequent decays will follow
with very short time delays related to nuclear reaction times, and possibly shorter then
the 10$^{-18}$ seconds of CN decay, actually a simultaneous decay can be considered. 
 This fact should be responsible for  the enhanced formation of resonances like $^{8}$Be and 
the $^{12}$C$^\star(0^{+}_2,2^{+}_2)$ states. An enhanced emission of multiple  
 $\alpha$-particles is predicted~\cite{vOe06}. Very relevant,
 however, is the larger radial extension of the Boson condensate states, as 
discussed in refs.~\cite{schuck04b,yamada04,Tza06a}.
\begin{figure}[htbp]
  \begin{center}
      \includegraphics[width=0.78\textwidth]{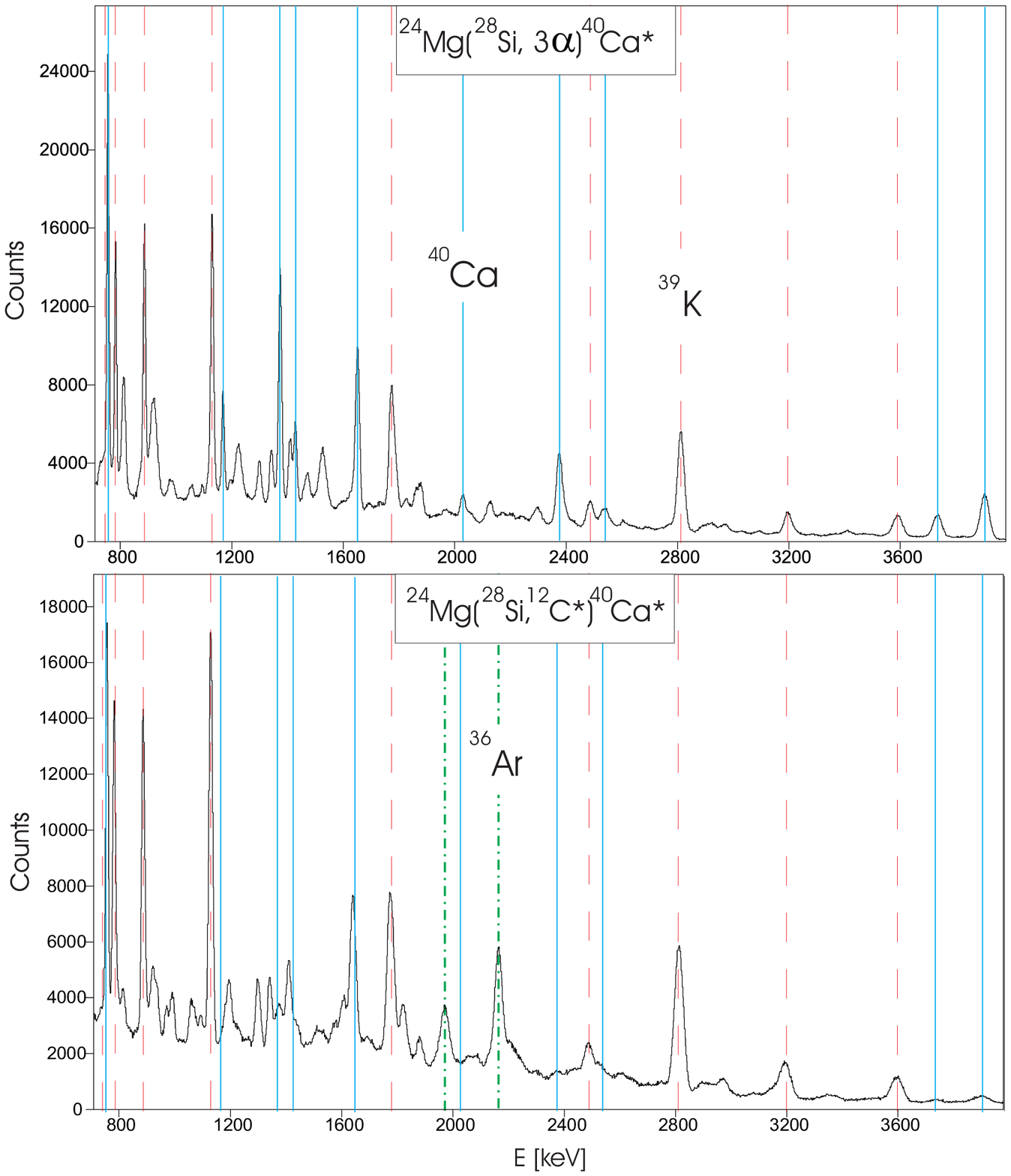}
     \caption {Coincident $\gamma$-spectra gated with the particles from 
               $\Delta$E-E-telescopes
              with the emission of three 
             random $\alpha$'s at different angles in different detectors (upper panel),  
            in comparison with that obtained by the $^{12}$C$^*_{0_+}$-gate (lower panel).
            The reaction is  $^{28}\textrm{Si}+^{24}\textrm{Mg} 
            \rightarrow ^{52}\textrm{Fe}
            \rightarrow ^{40}\textrm{Ca}+ 3\alpha$ at 130 MeV. Note the additional lines
             for $^{36}\textrm{Ar}$ in the lower panel  (courtesy of Tz. Kokalova). }
    \label{fig:gamc}
  \end{center}
\end{figure}

   The best way to study such decays is the combination of multi-detector arrays for 
particle detection with  $\Delta$E-E detectors and a ``calorimeter'' to observe 
the remaining compound nucleus residue via its $\gamma$-decay.
Such  experiments have been   performed with the large $\gamma$-detector array GASP at the 
Legnaro National Laboratory LNL at Padua (Italy), combined with the 
charged particle detector ball ISIS  (details are given in ref.~\cite{Tza05})
 consisting of 42 $\Delta$E-E telescopes (see Fig.~\ref{fig:gaspisis}). 

These experiments were performed in a study
of  $\gamma$-decays of compound nuclei selected with a particular 
particle decay~\cite{Torilov}.
The large  opening angle of the 
individual ISIS-$\Delta$E-E telescopes, which was 27$^{\circ}$, allows to select  
 the spontaneous decay of the weakly unbound states,
  namely of $^{8}$Be into two  $\alpha$'s and 
 the $^{12}$C$^*(0_{2}^+$), 
into three ${\alpha}$-particles. With the rather modest kinetic energy of these fragments and 
the small decay energies 
of  a few 100 KeV the opening angles between the  $\alpha$'s are in the 
range of 10$^{\circ}$ - 25$^{\circ}$,
 which fit into these solid angles. Therefore  
these prompt multiple $\alpha$-decays are observed by the pile-up of the signals
 produced by individual
 alpha-particles in one of the $\Delta$E-E telescopes. This is shown in Fig.~\ref{fig:dE_E_2}.
 The corresponding coincident 
(particle gated) $\gamma$-decays are compared with the spectra obtained from 
statistical emission (with the same  ${\alpha}$-multiplicity)
 into different $\Delta$E-E telescopes (see Fig.~\ref{fig:gamc}). 
\begin{figure*}[[htbp]
    \includegraphics[width=0.48\textwidth]{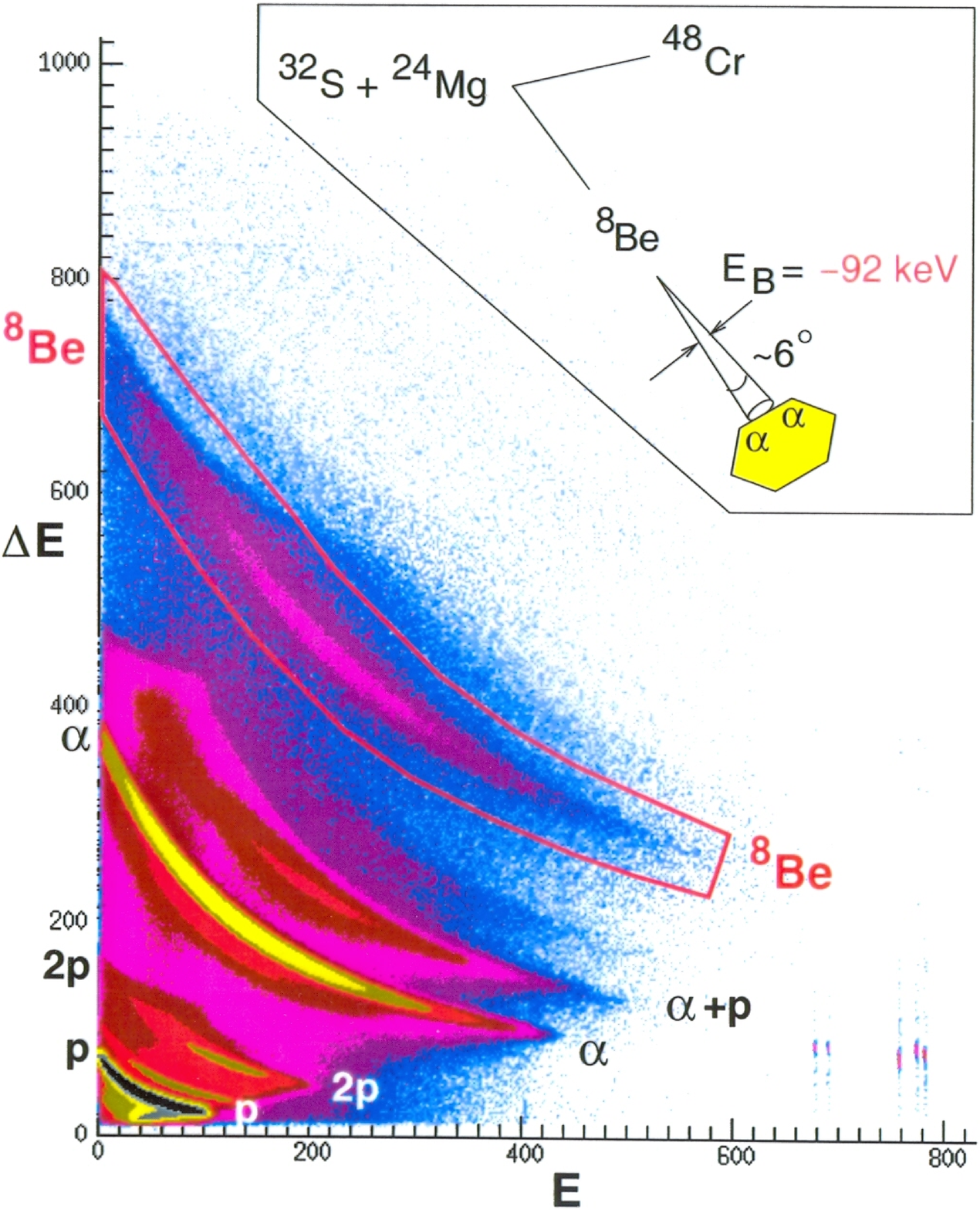}
 \hspace{2.0 mm} 
     \includegraphics[width=0.48\textwidth]{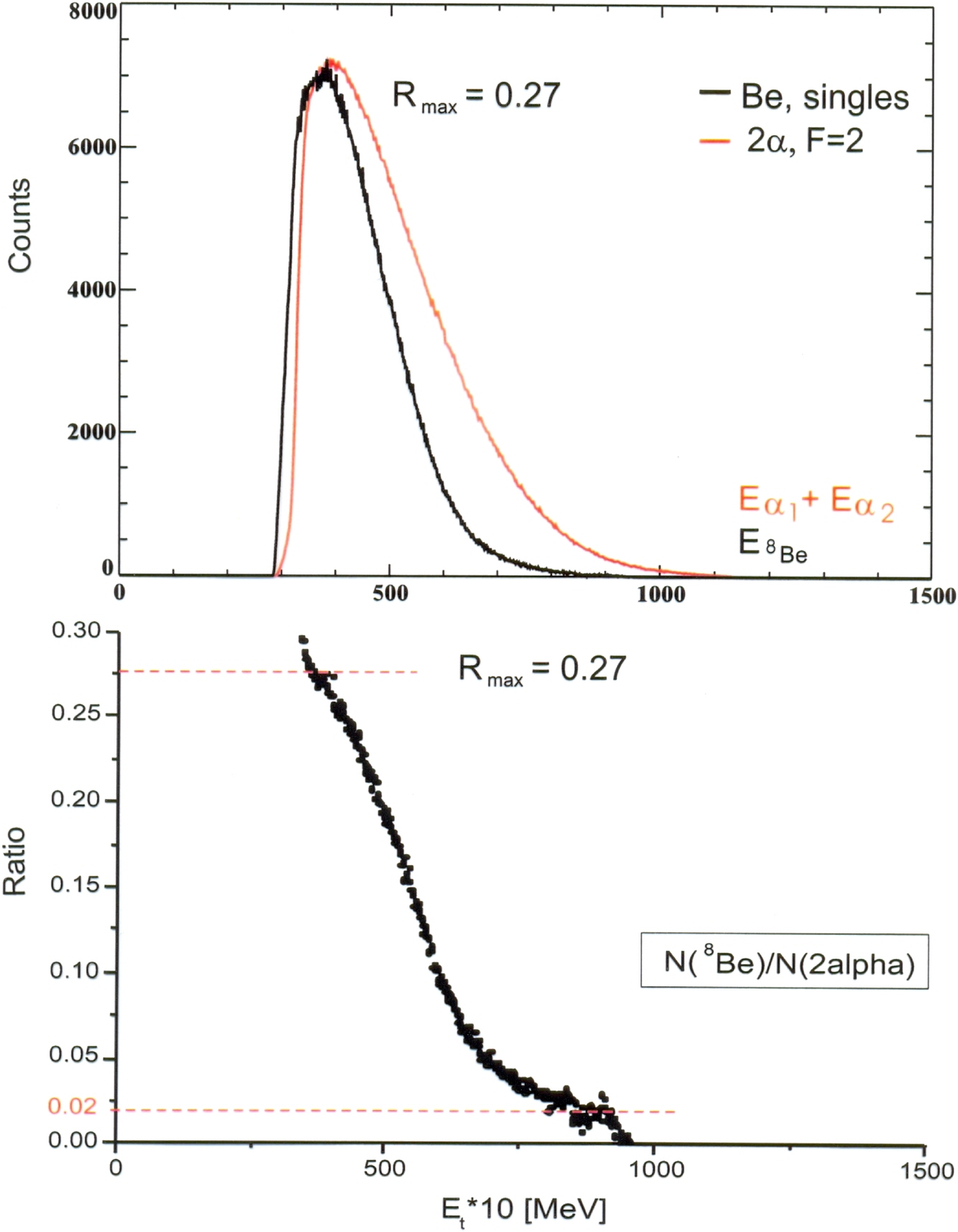} 
     \caption {The kinematics for  $^{8}$Be-observation, and comparison of the summed energies of 
               two  $\alpha$'s with the energy of  $^{8}$Be under the same kinematical conditions
                  (courtesy of S. Thummerer).}
    \label{fig:EBe8}
\end{figure*}

\begin{figure*}[[htbp]
  \begin{center}  
     \includegraphics[width=0.79\textwidth]{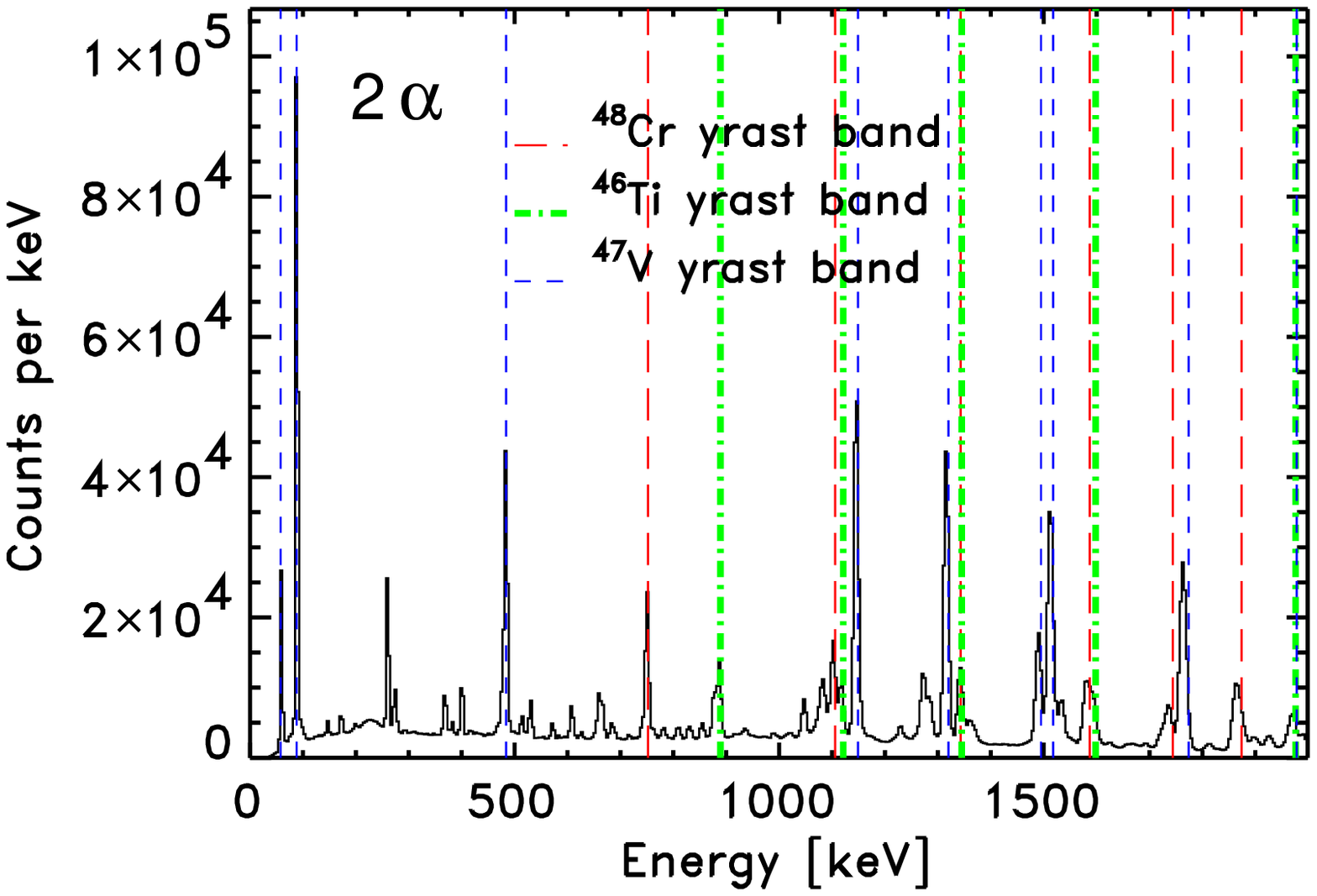} 
      \includegraphics[width=0.79\textwidth]{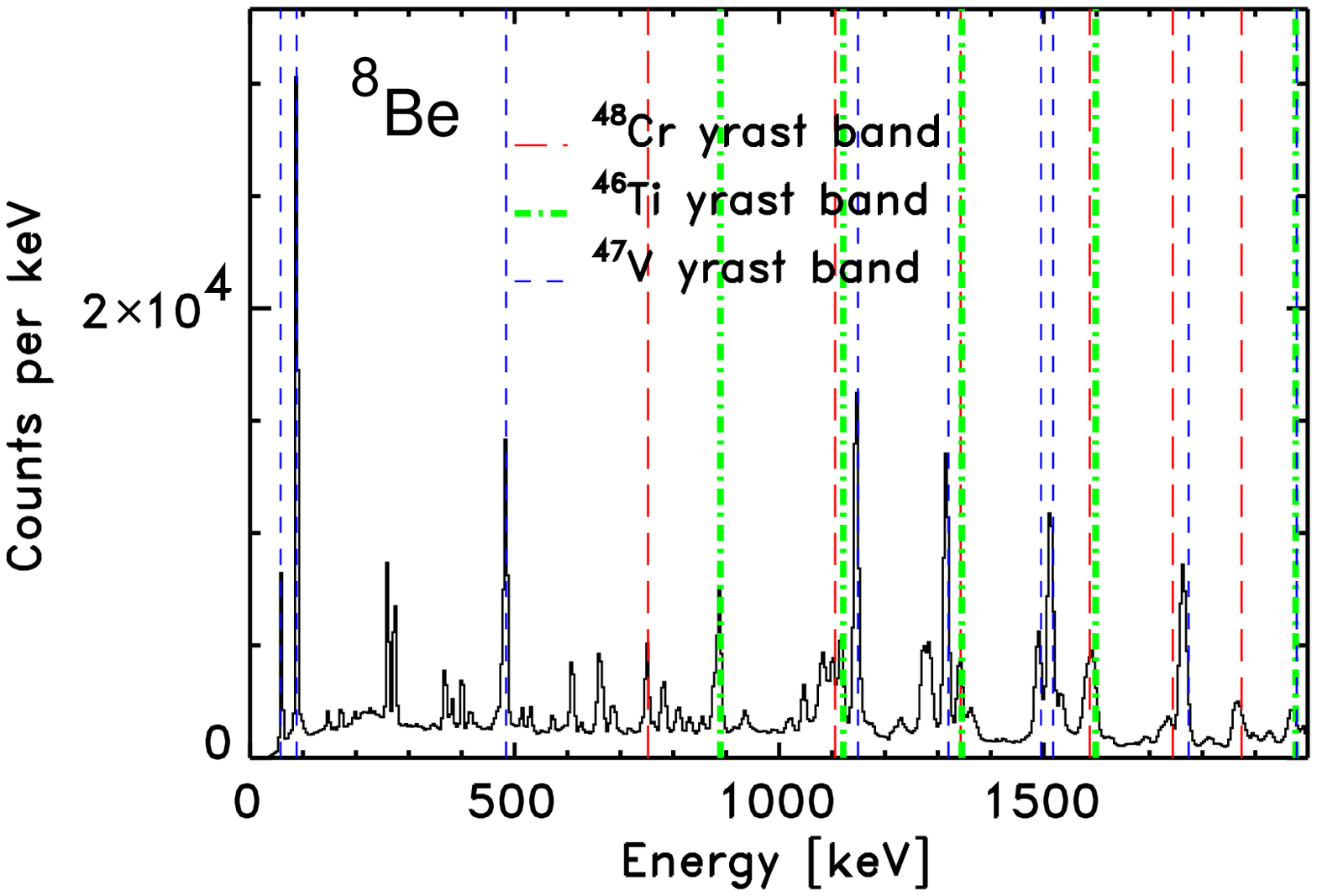}
     \caption {Coincident $\gamma$-spectra gated with the particles from 
               $\Delta$E-E-telescopes  
              with the emission of
            two  random $\alpha$'s in different detectors (upper panel),  
            in comparison with that obtained by the $^{8}$Be-gate (lower panel).
            The reaction is  $^{32}\textrm{S}~+~^{24}\textrm{Mg} 
            \rightarrow ^{56}\textrm{Ni}
            \rightarrow ^{48}\textrm{Cr}+ 2\alpha$ at 130 MeV  (courtesy of S. Thummerer).}
    \label{fig:gambe}
  \end{center}
\end{figure*}

The comparison of the two $\gamma$-sprectra with different triggers is
 shown in Fig.~\ref{fig:gamc} for the reaction 
 $^{28}\textrm{Si}~+~^{24}\textrm{Mg} 
            \rightarrow ^{52}\textrm{Fe} \rightarrow ^{40}\textrm{Ca} + 3\alpha$, 
an experiment designed 
for the spectroscopy of  $^{40}$Ca. The spectrum gated 
with three $\alpha$'s in one telescope shows additional  $\gamma$-transitions in 
  $^{36}$Ar, connected with an emission of 
an additional $\alpha$-particle, it is a dramatic effect, because these transitions 
are completely absent 
in the other spectrum gated by random directions of the 3  $\alpha$'s. Initial 
attempts to explain these differences by parameters of the statistical compound nucleus
decay failed, see ref.~\cite{Tza05}. A subsequent analysis~\cite{Tza06a}, which uses the 
the features of a  $\alpha$-condensed state, namely the larger diffuseness and the 
larger radial extension gave as an important effect a strong lowering 
by 10 MeV of the emission barrier
for the emission of  $^{12}$C$^*_{0^+}$. This fact explains, that the energies of
 the $^{12}$C$^*_{0^+}$ are concentrated at  much lower energies as compared to the summed
 energy of 3$\alpha$-particles
under the same kinematical conditions. In this way the residual nucleus  ($^{40}$Ca)
  attains a much higher residual
excitation energy~\cite{Tza05}.

 I also show the 
 results of  the previous study of the reactions  $^{32}\textrm{S}~+~^{24}\textrm{Mg}$
 for the  $\gamma$-spectroscopy of  $^{48}$Cr with 
 $^{8}$Be-emission~\cite{thummerer00,thummerer01,vOe00} performed with the same mentioned
ISIS-GASP-combinations at the LNL in Legnaro. In Fig.~\ref{fig:EBe8} we show the 
identification of  $^{8}$Be, and on the right side the comparison of the energy
spectra, under the same kinematical conditions, for  $^{8}$Be
 and the sum energy  of the two  $\alpha$'s. We note that the energy spectrum of the 
 $^{8}$Be is shifted to smaller energies, as in the previous case. This again must
 be explained by a larger diffuseness of the CN-state (an $\alpha$-condensed state)
 and a lowered Coulomb
 barrier for the  $^{8}$Be emission. The  $\gamma$-spectra with the 
two possible particle gates
  are shown in Fig.~\ref{fig:gambe}. It shows  
 the case of  $^{8}$Be-emission compared with the statistical emission of 
two  $\alpha$'s in two different detectors, the latter 
representing the usual statistically independent 
 decay into two different detection angles. 
Again we  found that the particular channel with  $^{8}$Be 
carries less energy and less angular momentum,  
therefore more subsequent decays are observed. In this case a subsequent 
neutron and proton emission is observed, the $^{46}$Ti-channel is strongly 
 increased for the $^{8}$Be-gate. 
Attempts to explain these differences in terms of parameters of CN-decay, (discussed
in ref.~\cite{thummerer01})
gave no conclusive result. At that time the concept of condensed  $\alpha$-particle states
in the CN was not considered.

\subsection{Inelastic excitation and fragmentation}
\label{exper:4}
The last entry in Tab.~\ref{tab:ecrit}
the last column for the $^{40}$Ca-core
has a negative sign  for $^{164}$Pb,  indicating that
this nucleus, as well as  lighter nuclei (actually above Z=72), are unstable 
in their ground state to single and multiple proton or multi-$\alpha$-particle  decay.
For lighter N=Z nuclei,  at excitation energies above E$^*_{crit}$
 another decay mode (as already mentioned) becomes  possible,
 which we call  Coulomb explosion. The condensed states are radial monopole excitations
 with respect to the ground state. The  monopole states located at 
high excitation energies can best be excited by Coulomb excitation at the highest projectile
 energies.
Coulomb excitations of the GQR (giant quadrupole resonance) or the GDR (giant dipole resonance)
 have been studied~\cite{Glasmacher}
 up to 300 MeV/Nucleon. The 
highest cross sections, also for the monopole excitations, 
are expected at projectile energies  above 1 GeV/nucleon. 
Because of the larger step in excitation energy the increment for the dynamical matching
becomes optimum at these highest energies. Such studies exist for some of the lighter
N=Z nuclei ($^{12}$C, 
$^{16}$O,$^{20}$Ne) and heavier~\cite{Zarubin}. In these studies nuclear emulsions 
have been used,
the silver nuclei (Ag) acting as target nuclei for Coulomb excitation.
\begin{figure}[htbp]
  \begin{center}
      \includegraphics[width=0.68\textwidth]{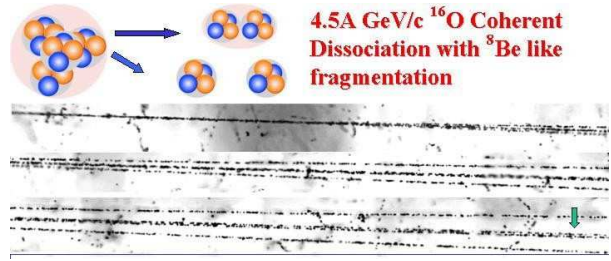}
     \caption {Break-up of  $^{16}\textrm{O}$ at 4.5 GeV/nucleon with the 
              emission of 4 
              $\alpha$'s, registered in an emulsion. Details of the decay can be seen,
              e.g. the more narrow cone of two $\alpha$'s,  due to
               the emission of $^8$Be. Different stages of   the
               decay, registered down stream in the emulsion are shown in consecutive
               panels.
               P. Zarubin private communication and ref.~\cite{Zarubin}}.
    \label{fig:16O-4a}
  \end{center}
\end{figure}

\begin{figure}[htbp]
  \begin{center}
      \includegraphics[width=0.68\textwidth]{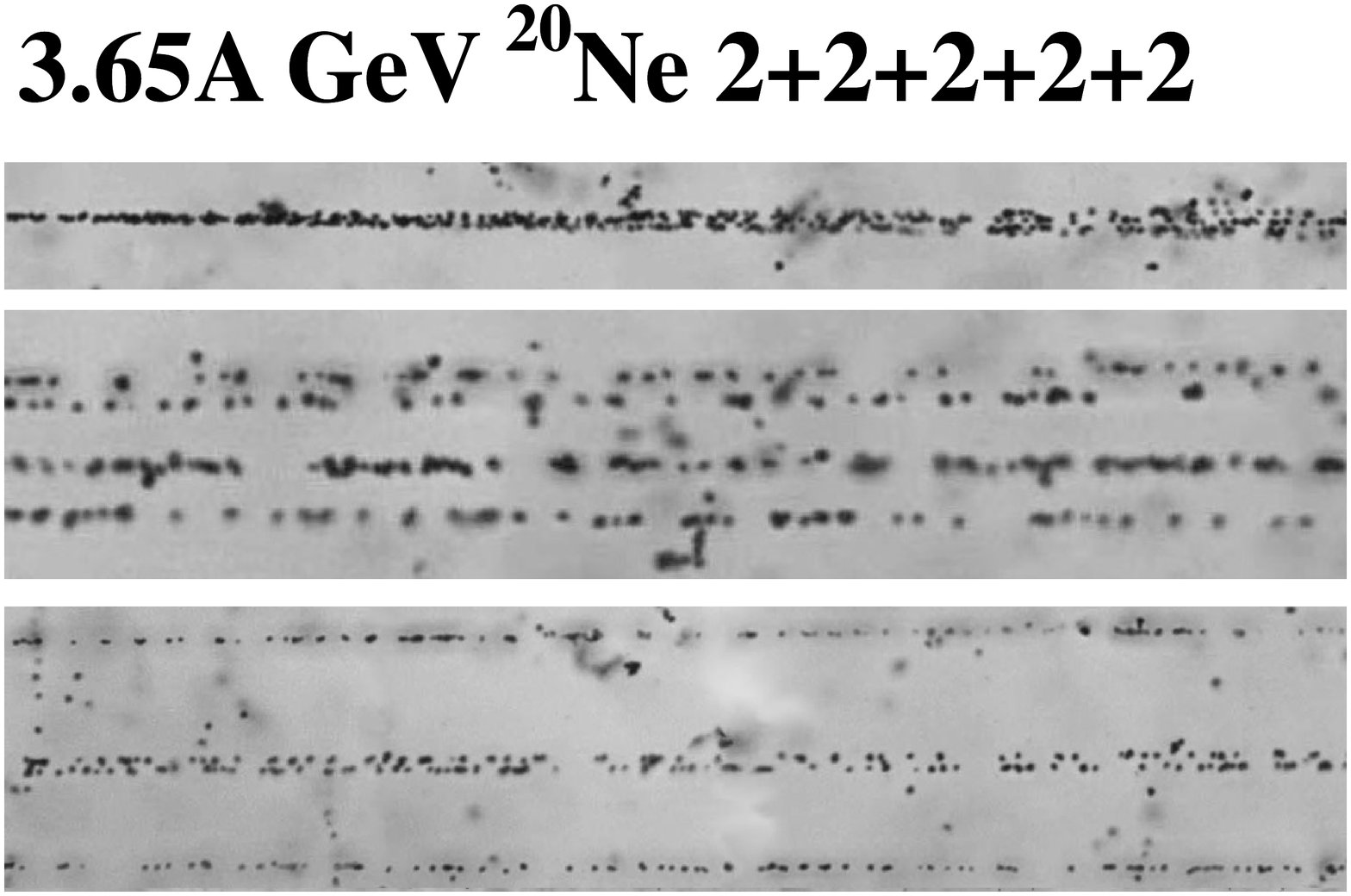}
     \caption {Break-up of  $^{20}\textrm{Ne}$ at 3.65 GeV/nucleon with the 
              emission of 5 
              $\alpha$'s (again partially as $^8$Be),
              registered in an emulsion. Different stages of the
               decay, registered down stream in the emulsion are shown in three panels
               on top of each other.
                P. Zarubin private communication and~\cite{Zarubin}}.
    \label{fig:Ne-5a}
  \end{center}
\end{figure}

The results were obtained at the JINR in Dubna with beams from
the Nucletron accelerator~\cite{Zarubin}. The reaction products are
 registered in nuclear emulsions, the Coulomb break-up being induced by the
heavy target nuclei (Silver, Ag) of the material. In this way very characteristic
multiple tracks after break-up have been observed (see Figs. \ref{fig:16O-4a}
and   \ref{fig:Ne-5a}). In the case of  $^{16}\textrm{O}$ we observe two  $\alpha$'s
and a  $^8$Be, this fact points to the previous discussions of a coherent emission,
two  $\alpha$'s must be emitted in a close correlation (in energy and space) 
in order to be able to form a  $^8$Be resonance.
 We expect the formation of $^8$Be from the internal structure of the condensate state
in  $^{16}\textrm{O}$, but also in the case of a simultaneous (coherent) emission
(and only in this case) the interaction of the two  $\alpha$'s can form  $^8$Be.

The result for the break-up of  $^{20}$Ne is shown in Fig.~\ref{fig:Ne-5a}, 
among the different observed break-up's the emission of   5-$\alpha$'s
 is observed with remarkable intensity.
Again at least one pair of $\alpha$'s is observed, indicating coherent emission 
with strong correlations, which 
allow the formation of the low lying  $^8$Be resonances.

In this context we mention that Coulomb explosion has been  observed in highly 
 charged atomic van der Waals clusters 
and is  discussed
by Last and Jortner~\cite{jortner}. In our case the simultaneous emission
of many $\alpha$-particles is expected, a decay process  very different from
 standard statistical compound nucleus decay. In fact in this decay mode the 
$\alpha$-particles must have all the same energy.

\subsection{ $\alpha$-$\alpha$ correlations}
 There have been numerous studies of particle-particle correlations in  higher
energy nuclear reactions around 50-100 MeV/Nucleon~\cite{pochod87}, 
as well as for reactions at relativistic energies, where
pion-pion correlations have been studied~\cite{wied99}. 
From this work we find that the spatial and time extension of the 
source can be studied in these correlations. A specific feature appears here,
that the correlations of bosons will exhibit a maximum at the smallest
 angles and smallest relative momenta. However, with two  $\alpha$-particles
the Coulomb interaction and the resonances in the $\alpha$-$\alpha$ channel, states in the 
$^{8}$Be nucleus dominate the correlations~\cite{pochod87}.
With an experimental set-up consisting of  $\Delta$E-E telescopes like the detector ball
EUROSIB  (a new development after ISIS described before)
 consisting of finer granulated  detectors, which contain sufficiently small angular
 resolution, and a $\gamma$-detector ball
 as in the experiments described in sect.~\ref{exper:3}, the 
 $\alpha$-$\alpha$ correlations
should be studied in coincidence with  $\gamma$-transitions of the 
residual N=Z compound nucleus (minus
two  $\alpha$-particles). With the use of inverted kinematics, the  
heavier projectiles on a lighter
target, the rather low energies of the  $\alpha$-particles in the cm system of 
the compound nucleus will have sufficiently
high energy in the laboratory system to be registered in  $\Delta$E-E-telescopes 
(an absorber has to be used to block the heavy 
projectiles).  A
  correlation matrix with ($E_{\alpha_1}$-$E_{\alpha_2}$), can be constructed.
 Such correlation matrices (for two  $\gamma$-rays) have been constructed in 
 $\gamma$-spectroscopy~\cite{stephens88} with 
 $\gamma$-detector balls. The correlations and the resonances can then be 
constructed over a wide range of relative and absolute momenta.

\section{Conclusions}
The data presented here show clearly experimental features, which point to the existence of 
$\alpha$-condensed states giving rise to  coherent multi $\alpha$-particle states 
in excited  N=Z nuclei. With the most
 recent experimental developments, we can expect that important new features of such
 states can be observed. These can potentially establish the existence of Bose-Einstein
condensates in nuclei, a very promising field of research for future studies.
\begin{acknowledgement} 
The author is indebted to Peter Schuck for numerous discussions.
\end{acknowledgement}


\begin{thebibliography}{99}
\bibitem{bm}        Aage~Bohr and Ben~R.~Mottelson, {\em Nuclear Structure},
                    Vol.~{\bf I} (W.A. Benjamin Inc., 1975).

\bibitem{Feldmeier95}  H.~Feldmeier, K.~Bieler and J.~Schnack,
                    Nucl. Phys. A {\bf 586}, 493 (1995).

\bibitem{Neff05}    T.~Neff, H.~Feldmeier and R.~Roth,  
                    Nucl. Phys. A {\bf 752}, 94 (2005). 

\bibitem{horiuchi86}  H.~Horiuchi and K.~Ikeda, Cluster model of the 
                    Nucleus, International Review of Nuclear Physics 
                    Vol.~{\bf 4} (World Scientific, Singapore, 1986) p.1
                    and references therein.

\bibitem{Hor97}     H. Horiuchi and Y. Kanada-En'yo,
                    Nucl. Phys. A {\bf 616}, 394c (1997).



\bibitem{Kan01}     Y.~Kanada-En'yo and H.~Horiuchi,
                    Prog. Theor. Phys., Suppl. No. {\bf 142}, 206 (2001).



\bibitem{voe06}     W.~von~Oertzen, M. Freer and Y. Kanada-Enyo,
                    Phys. Reports {\bf 432}, 43 (2006). 


\bibitem{horiuchi68} H.~Horiuchi, K.~Ikeda, Prog. Theor. Phys. (Japan)
                     A {\bf 40}, 277 (1968), and
                     K.~Ikeda, N.~Tagikawa, H.~Horiuchi, Prog. Theor. Phys.
                    (Japan), Extra Vol. Suppl. 464 (1968).

\bibitem{Horiuchi70} H.~Horiuchi, Prog. Theor. Phys. {\bf 43}, 375 (1970).

\bibitem{Ikeda72}   H.~Horiuchi, K.~Ikeda, Y.~Suzuki, Prog. Theor. Phys.
                    (Japan) Suppl. {\bf 52} Chapt.3 (1972).

\bibitem{vOe06}     W. von Oertzen, 
                    Eur. Phys. J. A {\bf 29}, 133 (2006).

\bibitem{audi03}    G. Audi et al.,  Nucl. Phys. A {\bf 729}, 3-128 (2003). 
                   
\bibitem{isakov00}  V. I. Isakov et al., Particles and Nuclei Letters {\bf 5}, 44
                       (2000). 

\bibitem{hoyle}     F. Hoyle, The Atrophys. J. Suppl. Ser. {\bf 1},12 (1954).                 

\bibitem{tohsaki01} A.~Tohsaki, H.~Horiuchi, P.~Schuck and G.~R\"{o}pke,
                    Phys. Rev. Lett. {\bf 87}, 192501 (2001).

\bibitem{schuck04a} P.~Schuck, Y.~Funaki, H.~Horiuchi, G,~R\"{o}pke, A.~Tohsaki 
                    and T.~Yamada, Nucl. Phys. A {\bf 738}, 94 (2004).

\bibitem{schuck04b} Y.~Funaki, H.~Horiuchi, W von Oertzen  {\em et al.}, 
                    Phys. Rev. C {\bf 80}, 064326 (2009).                   

\bibitem{yamada04}  T.~Yamada and P.~Schuck,
                    Phys. Rev. C {\bf 69}, 024309 (2004).

\bibitem{Schuck07}  P. Schuck, Y. Funaki, H. Horiuchi, G. R\"opke, A. Tohsaki,
                    T. Yamada,
                    Progress in Particle and Nuclear Physics{\bf 59} 285 (2007)  und idem,
                    Nuc. Phys. News {\bf 17} 11 (2007). 

\bibitem{funaki09}  Y. Funaki {\em et al.}, Phys. Rev. {\bf 80}, 064326 (2009)

\bibitem{freer07}    M. Freer, Rep. Prog. Phys. {\bf 70}, 2149 (2007).

\bibitem{amit}      D. J.~Amit and Y.~Verbin, Statistical Physics, 
                    An Introductory Course (World Scientific, 1995).


\bibitem{FH}        H.~Frauenfelder and E.M.~Henley, Subatomic Physics, p.546.
                  Prentice Hall Inc., Physics Series (1974),
                  Englewood Cliffs (New Jersey).


\bibitem{Torilov}    S. Torilov  {\em et al.}, Eur. Phys. J. A {\bf 19}, 307
                    (2004).

\bibitem{Tza05}     Tz.~Kokalova {\em et al.}, Eur. Phys. J. A {\bf 23}, 19
                    (2005).

\bibitem{Tza06a}    Tz.~Kokalova, N.~Itagaki {\em et al.}, Phys. Rev. Lett.,
                    {\bf 96}, 192502 (2006). 

\bibitem{thummerer00} S.~Thummerer {\em et al.}, 
                      Phys. Scr. \textbf{T~88},  114 (2000).

\bibitem{thummerer01} S.~Thummerer, W. von~Oertzen, B.~Gebauer, 
                      S.~Lenzi, A.~Gadea, D.R.~Napoli, 
                      C.~Beck and M.~Rousseau, J. Phys. \textbf{G 27}, 1405 (2001).
\bibitem{vOe00}
                      W.~von~Oertzen, Phys. Scr. T \textbf{88}, 83 (2000).

\bibitem{Khoa07}     Dao T. Khoa, W. von Oertzen, H. G. Bohlen and S. Ohkubo
                     J. Phys. G; Nucl. Part. Phys. {\bf 34}, R111-R164
                    (2007).

\bibitem{Chern07}    M. Chernych {\em et al.}, Phys. Rev. Letters {\bf 98}, 032501 (2007).

\bibitem{Khoa08}     Dao T. Khoa and Do Cong Cuong, Phys. Lett. B {\bf 660}, 331 (2008)
                       and private communication.

\bibitem{Danilov09}  A. N. Danilov  {\em et al.}    Phys. Rev. C {\bf 80}, 054603 (2009).

\bibitem{Wakasa07}   T. Wakasa {\em et al.} Phys. Letters B {\bf 653}, 632 (2007)



\bibitem{broglia}   R.~Broglia, Ann. Phys. (New York), {\bf 80} 60 (1973). 

\bibitem{voevitt01} W.~von~Oertzen and A.~Vitturi, 
                    Rep. Prog. Phys. {\bf 64}, 1247 (2001).

\bibitem{voe76}     W.~von~Oertzen, Phys. Lett. {\bf 61B}, 223 (1976). 

\bibitem{Glasmacher} T. Glasmacher, Annu. Rev. Nucl. Part. Sci.{\bf 48}, 1 (1998).  

\bibitem{jortner}   I.~Last and J.~Jortner, 
                    Phys. Rev. A {\bf 62}, 013201 (2000) and references therein.

\bibitem{Zhereb07}  V. Zherebchevsky {\em et al.}, 
                    Phys. Letters. {\bf 646 B}, 12 (2007)

\bibitem{VOe08}     W.~von~Oertzen {\em et al.}, Eur. Phys. J. A {\bf 36}, 279 (2008) 


\bibitem{Zarubin}  V. Bradnova {\em et al.}, Nucl. Phys. A {\bf 734}, E92 (2004) 
                    and P. Zarubin private communication (2009).


\bibitem{pochod87}  J. Pohodzalla {\em et al.},
                    Phys. Rev. C {\bf 35}, 1695 (1987).

\bibitem{wied99}    U. A. Wiedemann and U. Heinz, 
                    Phys. Rep. {\bf 319}, 145 (1999).

\bibitem{stephens88}  F. S. Stephens {\em et al.},
                    Phys. Rev. Lett.  {\bf 37}, 2927 (1988).


\end{thebibliography}
\end{document}